\documentclass[twocolumn]{aastex631}

\usepackage[separate-uncertainty = true,multi-part-units=single,range-units=single]{siunitx}
\usepackage{soul} 

\received{April 5, 2023}
\revised{September 12, 2023}
\accepted{September 12, 2023}

\shorttitle{Switchback boundaries}
\shortauthors{Bizien et al.}

\begin{document}

\title{Are Switchback boundaries observed by Parker Solar Probe closed?}

\correspondingauthor{Nina Bizien}
\email{nina.bizien@cnrs-orleans.fr}

\author[0000-0001-6767-0672]{Nina Bizien}
\affiliation{LPC2E, CNRS/University of Orléans/CNES, Orléans, France}

\author[0000-0002-4401-0943]{Thierry Dudok de Wit}
\affiliation{LPC2E, CNRS/University of Orléans/CNES, Orléans, France}
\affiliation{International Space Science Institute, 3012, Bern, Switzerland}

\author[0000-0001-5315-2890]{Clara Froment}
\affiliation{LPC2E, CNRS/University of Orléans/CNES, Orléans, France}

\author[0000-0002-2381-3106]{Marco Velli}
\affiliation{Department of Earth, Planetary, and Space Sciences, University of California, Los Angeles, CA 90095, USA}

\author[0000-0002-3520-4041]{Anthony W. Case}
\affiliation{BWX Technologies, Inc., Lynchburg VA 24501, USA}

\author[0000-0002-1989-3596]{Stuart D. Bale}
\affiliation{Physics Department, University of California, Berkeley, CA 94720-7300, USA}
\affiliation{Space Sciences Laboratory, University of California, Berkeley, CA 94720-7450, USA}

\author[0000-0002-7077-930X]{Justin Kasper}
\affiliation{BWX Technologies, Inc., Washington DC 20002, USA}
\affiliation{Climate and Space Sciences and Engineering, University of Michigan, Ann Arbor, MI 48109, USA}

\author[0000-0002-7287-5098]{Phyllis Whittlesey}
\affiliation{Space Sciences Laboratory, University of California, Berkeley, CA 94720-7450, USA}

\author[0000-0003-3112-4201]{Robert MacDowall}
\affiliation{Solar System Exploration Division, NASA/Goddard Space Flight Center, Greenbelt, MD, 20771, USA}

\author[0000-0001-5030-6030]{Davin Larson}
\affiliation{Space Sciences Laboratory, University of California, Berkeley, CA 94720-7450, USA}

\begin{abstract}

Switchbacks are sudden and large deflections in the magnetic field that Parker Solar Probe frequently observes in the inner heliosphere. Their ubiquitous occurrence has prompted numerous studies to determine their nature and origin. Our goal is to describe the boundary of these switchbacks using a series of events detected during the spacecraft's first encounter with the Sun. Using FIELDS and SWEAP data, we investigate different methods for determining the boundary normal. The observed boundaries are arc-polarized structures with a rotation that is always contained in a plane. Classical minimum variance analysis (MVA) gives misleading results and overestimates the number of rotational discontinuities. We propose a robust geometric method to identify the nature of these discontinuities, which involves determining whether or not the plane that contains them also includes the origin ($\textbf{\textit{B}}=0$). Most boundaries appear to have the same characteristics as tangential discontinuities in the context of switchbacks, with little evidence for having rotational discontinuities. We find no effect of the size of the Parker spiral deviation. Furthermore, the thickness of the boundary is within MHD scales. We conclude that most of the switchback boundaries observed by Parker Solar Probe are likely to be closed, in contrast to previous studies. Our results suggest that their erosion may be much slower than expected. 
\end{abstract}

\keywords{Solar wind; Interplanetary magnetic fields, Interplanetary discontinuities}

\section{Introduction} \label{sec_introduction}

Sudden magnetic deflections in the inner heliospheric solar wind, called switchbacks, have recently attracted considerable attention. These deflections from the Parker spiral are best observed by the Parker Solar Probe \citep[PSP;][]{foxSolarProbeMission2016}. Their origin and nature are still unestablished. They are considered to contribute to the heating of the solar wind and acceleration of particles, as their omnipresence is observed in the young solar wind \citep{baleHighlyStructuredSlow2019, kasperAlfvenicVelocitySpikes2019, dudokdewitSwitchbacksNearSunMagnetic2020, horburySharpAlfvenicImpulses2020}. 
Similar structures have been observed farther away from the Sun by missions such as Ulysses, Helios, and ACE. However, close to the Sun, they are much more frequent and abrupt.

Numerous studies have sought to characterize switchback structures and understand their influence on the solar wind.  It has been shown that the switchbacks tend to appear in patches and that the size scale is about the same order as the supergranulation scale \citep{baleSolarSourceAlfvenic2021,fargetteCharacteristicScalesMagnetic2021}. Besides the consequent spatial scale, a temporal scale is observed during periods of corotation \citep{shiPatchesMagneticSwitchbacks2022}. Switchbacks appear to have a preferential orientation, in the clockwise tangential direction \citep{fargettePreferentialOrientationMagnetic2022,lakerSwitchbackDeflectionsEarly2022}.
The pitch angle of the strahl electrons does not change during the crossing of the structures, revealing that these are not heliospheric current sheet (HCS) crossings but rather bendings of the magnetic field away from the Parker spiral, and sometimes to the Sun \citep{kasperAlfvenicVelocitySpikes2019, whittleseySolarProbeANalyzers2020}. 

Switchbacks carry the signature of Alfvén waves because their magnetic intensity is nearly constant across their boundary. In addition, changes in magnetic field and velocity field are correlated \citep{kasperAlfvenicVelocitySpikes2019}. However, they are not purely Alfvénic since they are also weakly compressible, and dropouts of $|\textbf{\textit{B}}|$ have been observed at their boundary \citep{farrellMagneticFieldDropouts2020, larosaSwitchbacksStatisticalProperties2021}. For that reason, they are sometimes referred to as near Alfvénic. Many switchbacks present spikes in the velocity field that are due to the Alfvénicity   \citep{kasperAlfvenicVelocitySpikes2019}. There is also evidence for magnetic reconnection at the boundary of some switchbacks, yet this appears to be infrequent \citep{fromentDirectEvidenceMagnetic2021}. Finally, wave activity is enhanced at their boundary \citep{krasnoselskikhLocalizedMagneticfieldStructures2020, malaspinaInhomogeneousKineticAlfven2022} and \cite{dudokdewitSwitchbacksNearSunMagnetic2020} have shown there is no characteristic magnitude of the deflection of switchbacks, but that smaller ones are much more numerous.  

Despite the constraints brought by all these properties, many questions remain open as to the origin of switchbacks and their evolution in the corona. In particular, their evolution should be linked to the nature of their boundaries. Are these open with a possible plasma flow across them, or do they prevent any plasma flow, acting as closed boundaries? Similarly, the generation mechanisms of switchbacks may be constrained by the nature of the boundaries.

Two classes of mechanisms have been proposed for switchback generation. The first one is rooted deep in the solar atmosphere and involves their generation by interchange reconnection \citep{fiskGlobalCirculationOpen2020, drakeSwitchbacksSignaturesMagnetic2021, schwadronSwitchbacksExplainedSuperParker2021}. The other class is based on a local origin, directly in the solar wind. The theories revolve around in-situ phenomena, such as the turbulent evolution of Alfvén waves expanding in the solar wind \citep{squireInsituSwitchbackFormation2020, squirePropertiesAlfvenicSwitchbacks2022}, or the instability of naturally present shear flows in the solar wind \citep{ruffoloSheardrivenTransitionIsotropically2020}. These two classes and the associated theories are not mutually exclusive.

Switchbacks have been shown to evolve as they propagate \citep{teneraniEvolutionSwitchbacksInner2021}. However, it is still unclear whether their occurrence decreases when moving away from the Sun, as this depends very much on how these structures are defined. A common conclusion is that they exist everywhere in the solar wind, from the areas closest to the sun to $1~\si{\astronomicalunit}$, and beyond. Furthermore, their occurrence appears to depend on the type of sources from which they originate \citep{fargetteCharacteristicScalesMagnetic2021}.

We propose to focus here on the switchback boundaries. We will show that they have very specific properties, which questions the way boundaries are routinely analyzed in the solar wind. Our approach aims to answer the following open questions: 
\begin{itemize}
\item What is the nature of switchback boundaries?
\item How are the properties of the switchback boundaries linked to their dynamical evolution in the solar wind?
\end{itemize} 
 In Section \ref{sec_data}, we introduce the data we use and detail the identification of our set of boundaries. In Section \ref{sec_methods}, we analyze the characteristics of switchback boundaries and the limitations of the usual technique, the minimum variance analysis (MVA) method. We propose a different methodology to investigate their nature. In Section \ref{sec_results}, we present the switchback boundary properties retrieved from the newly implemented methodology. We conclude in Section \ref{sec_conclusion}. 
\section{Data}\label{sec_data}

We primarily use DC magnetic field measurements from the fluxgate magnetometer (MAG), which is part of the FIELDS suite \citep{baleFIELDSInstrumentSuite2016}. The distribution of the strahl electron pitch angle comes from the SPAN-E instrument \citep{whittleseySolarProbeANalyzers2020}, and the proton velocity is derived for the ion distribution function measured by the Faraday cup \citep[SPC;][]{caseSolarProbeCup2020}. Both particle instruments are part of the SWEAP suite \citep{kasperSolarWindElectrons2016}.

Throughout the paper, we use the RTN coordinate frame: the radial \textit{R} vector is along the Sun-spacecraft direction, taken anti-sunward; the \textit{T} component is the tangential vector, the result of the cross product of the solar rotation vector with \textit{R}; and the normal \textit{N} component, which completes the right-handed triad.
We identified 125 structures in the first PSP encounter from a comprehensive catalog of switchbacks.
We use a normalized deflection from the Parker spiral \citep{dudokdewitSwitchbacksNearSunMagnetic2020} to manually identify each switchback boundary in MAG data:
\begin{equation}\label{eq_deflection}
z=\frac{1}{2}(1-\cos~\alpha),
\end{equation} 
where $\alpha$ is the deflection angle between the measured magnetic field vector and the direction of the Parker spiral. The Parker spiral is identified in the data using quiet intervals with no switchbacks, and small normal and tangential components.  The deflections we selected in the catalog have a magnitude of $z \geq 0.1$, which corresponds to an angular deflection of  $\ge 26 \si{\degree}$. We noticed a clear modification of the statistical properties of the deflections below $z \approx 0.1$, which are increasingly difficult to distinguish from stochastic fluctuations. Let us stress that only large ones with $z>0.5$ (corresponding to a deflection of $> 90 \si{\degree}$) strictly qualify as switchbacks. However, for the sake of simplicity, all deflections with $z>0.1$ will be called switchbacks in what follows. The impact of the deflection on the properties of the boundary will be addressed later.

In the preprocessing, we also removed short-lived events lasting less than $3~\si{\s}$ and those whose change in the orientation of the magnetic field is not abrupt enough. Indeed, these are too difficult to distinguish from stochastic fluctuations. The deflections were then validated individually, taking into account the strahl pitch angle, to remove crossings of the heliospheric current sheet, or other artifacts.

\begin{figure*}
\centering	
\includegraphics[width=17cm]{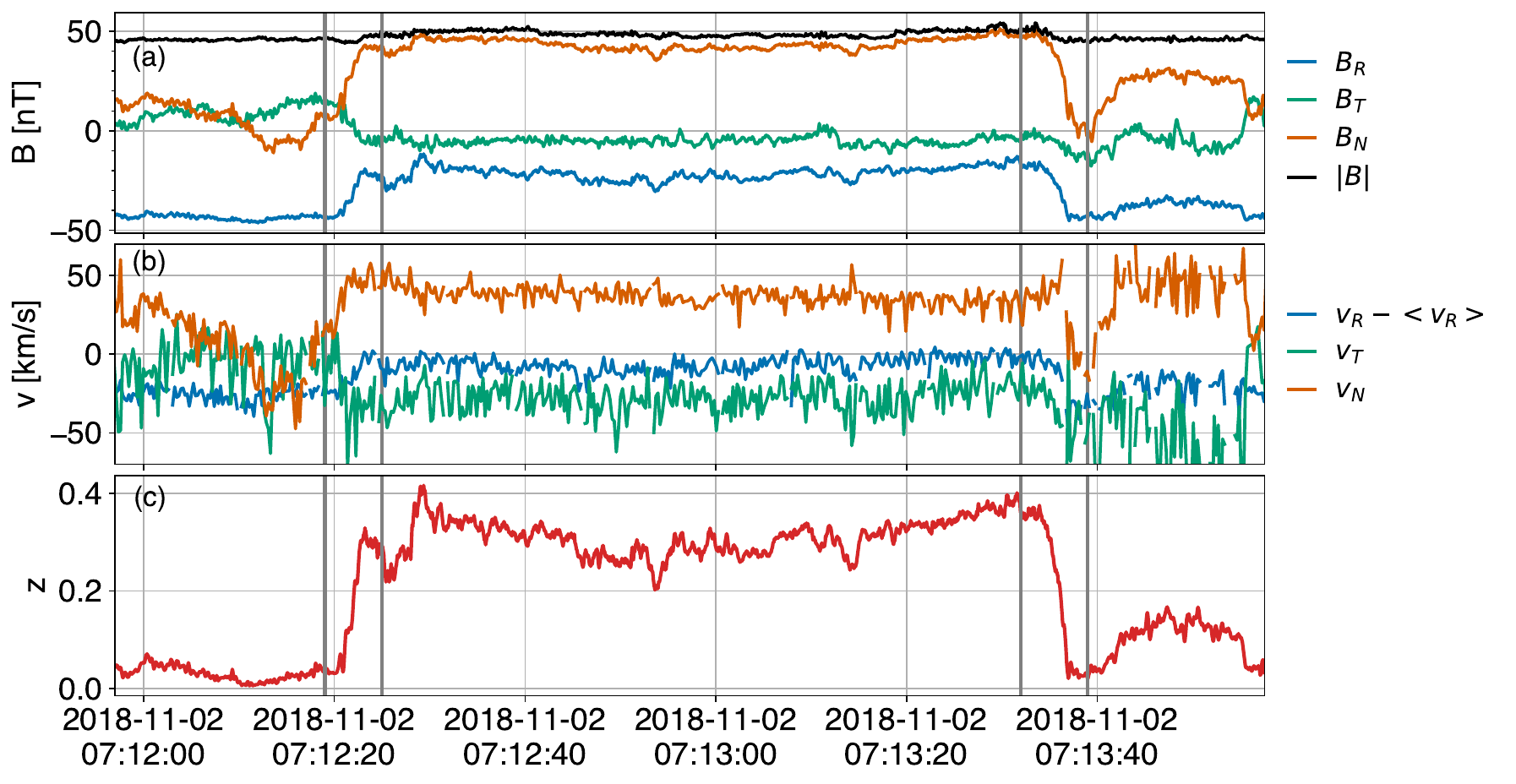}
 \caption{Example of a typical switchback. The leading and trailing boundaries are located between the pairs of vertical lines. (a) FIELDS/MAG magnetic field data in the RTN frame; (b) proton velocity from SWEAP/SPC, also in the RTN frame. The average value of the radial velocity has been removed to facilitate visualization; (c) normalized deflection $z$.}
\label{fig_ex_sb}
\end{figure*}

In Figure \ref{fig_ex_sb}, we present a typical switchback with well-defined boundaries. The boundaries are sharp and do not contain embedded substructures. The field moves from a quiet region (before) to another region (during) forming a sharp transition, which we define as the switchback boundary. Then, the field returns to another quiet region \textit{After} with another sharp boundary. The evolution of the proton velocity shows a high correlation with the magnetic field, which is typical for Alfvénic structures and has been observed for many switchbacks \citep{kasperAlfvenicVelocitySpikes2019,horburySharpAlfvenicImpulses2020}. We characterize the magnitude of the deflection by using the $\mathit{z}$-parameter. In this example, the maximal deflection with respect to the Parker spiral is approximately $66 \si{\degree}$.                                                                                                                                              

\section{Methods}\label{sec_methods}

\subsection{Context of Switchback boundaries}
The boundaries of switchbacks are characterized by an abrupt change in the magnetic field. In the MHD approximation, such discontinuities in the solar wind are routinely described by means of the Rankine-Hugoniot relations, whose main solutions are rotational discontinuities (RDs) and tangential discontinuities \citep[TDs;][]{hudsonDiscontinuitiesAnisotropicPlasma1970}. The latter are convected structures with closed boundaries that do not allow for particle transport across the discontinuity. On the contrary, RDs are isentropic, propagate with the Alfvén velocity, and allow for particle transport. Both types of discontinuities have received considerable attention in the solar wind because of the implications they have on its dynamical evolution.

A notable difference between TDs and RDs is the existence of a normal component of the magnetic field to the plane of the discontinuity. This normal, together with the properties of the solar wind will be our main parameters for helping classify the switchback boundaries and understanding their nature.

\subsection{Definition of a Normal vector}

We can define the normal to the discontinuity as a vector that is perpendicular to the field vector throughout the transition
\begin{equation}\label{eq_normal_scalar_product}
\mathbf{B}_{b} \cdot \mathbf{n}= \mathbf{B}_{a} \cdot \mathbf{n}=\mathbf{B}_{d} \cdot \mathbf{n},
\end{equation} 
where $\textit{\textbf{n}}$ is the normal to the plane in which the tips of the magnetic field vector rotate; the subscripts a (after) and b (before)  correspond to field vectors taken on each side of the discontinuity, and the subscript d (during) refers to inside the discontinuity \citep{sonnerupMagnetopauseStructureAttitude1967}.
We can then obtain the normal $\textit{\textbf{n}}$ by using the cross product of two vectors that are tangential to the discontinuity
\begin{equation}\label{eq_normal}
\mathbf{n}=\pm (\mathbf{B}_{b}-\mathbf{B}_{a}) \times (\mathbf{B}_{b}-\mathbf{B}_{d}),
\end{equation}

Knowing this normal and then projecting the measured field onto it formally allows us to distinguish closed and open discontinuities. TDs have a small scalar product $(\textbf{\textit{B}}/|\textbf{\textit{B}}|) \cdot \textbf{\textit{n}}$ while for RDs, $(\textbf{\textit{B}}/|\textbf{\textit{B}}|) \cdot \textbf{\textit{n}}$ is large.

Although this concept, which is based on three vectors only, is conceptually simple, it is not robust when the magnetic field is affected by turbulence or by wave activity. Furthermore, it is preferable to make use of all the available data and not just the three values of $\textbf{\textit{B}}$.

With multi-spacecraft data, finding the normal can be done by reconstructing the shape of the discontinuity in 3D \citep{knetterDiscontinuityObservationsCluster2003}. With single-spacecraft data, as is the case here, reconstructing the normal is usually done by means of MVA \citep{sonnerupMinimumMaximumVariance1998}.  While the application of the MVA method is in principle straightforward, the interpretation of its results comes with several challenges \citep{hausmanDeterminingNatureOrientation2004}. 

\subsection{MVA analysis}
The MVA method was designed to find the direction in space along which the scalar product $\textbf{\textit{B}} \cdot \textbf{\textit{n}}$ has the minimum variance over the interval of interest. The method does so by approximating the orbit of $\textbf{\textit{B}}$ (i.e. the locus of the tips of the magnetic field) by an ellipsoid. This ellipsoid can be conveniently described by its three orthogonal (called principal) axes and the variance $\lambda_i$ of the projection of $\textbf{\textit{B}}$ along each of these axes. By convention, these axes are defined so that $\lambda_1 \ge \lambda_2 \ge \lambda_3 \ge 0$ \citep[see][]{sonnerupMinimumMaximumVariance1998}.

The relative magnitudes of $\lambda_i$ lead to different shapes of the ellipsoid. An interesting case arises when there is a $\lambda_1 \ge \lambda_2 \gg \lambda_3$ degeneracy, as this corresponds to a 2D or 1D distribution of the orbit of $\textbf{\textit{B}}$. This is typical for plane-polarized waves, but also occurs with some types of discontinuities. In such circumstances, the third principal axis, along which the variance is smallest, is usually interpreted as the direction of the normal to the 2D structure. This is indeed the case for most of the discontinuities to be discussed below.

The limitations and misinterpretations of MVA methods have been extensively discussed in the literature \citep{knetter05,tsurutaniCommentCommentAbundances2007,tehLocalStructureDirectional2011,rosa_oliveira20}. Much confusion results from excessive reliance on the interpretation of the eigenvalues $\lambda_i$, which are a necessary but not sufficient condition for the interpretation of the results \citep{knetter05}. 

Regarding discontinuity analysis, a major limitation comes from the mathematical definition of the principal axes and $\lambda_i$.  Both are obtained by diagonalizing the covariance matrix of the magnetic field, which requires the prior subtraction of the mean of each of the components of $\textbf{\textit{B}}$. For most switchbacks, this mean depends strongly on the choice of the time interval around the discontinuity from which the mean is estimated. This arbitrariness in turn affects the value of the mean and the orientation of the normal \citep{hausmanDeterminingNatureOrientation2004}.

For these reasons, and as with all statistical analyses, it is vital to take a step back, and consider the raw data before interpreting the results of the MVA method. This is what we shall do next.

\subsection{A typical example studied using MVA}

\begin{figure*}
\centering	
\includegraphics[width=17cm]{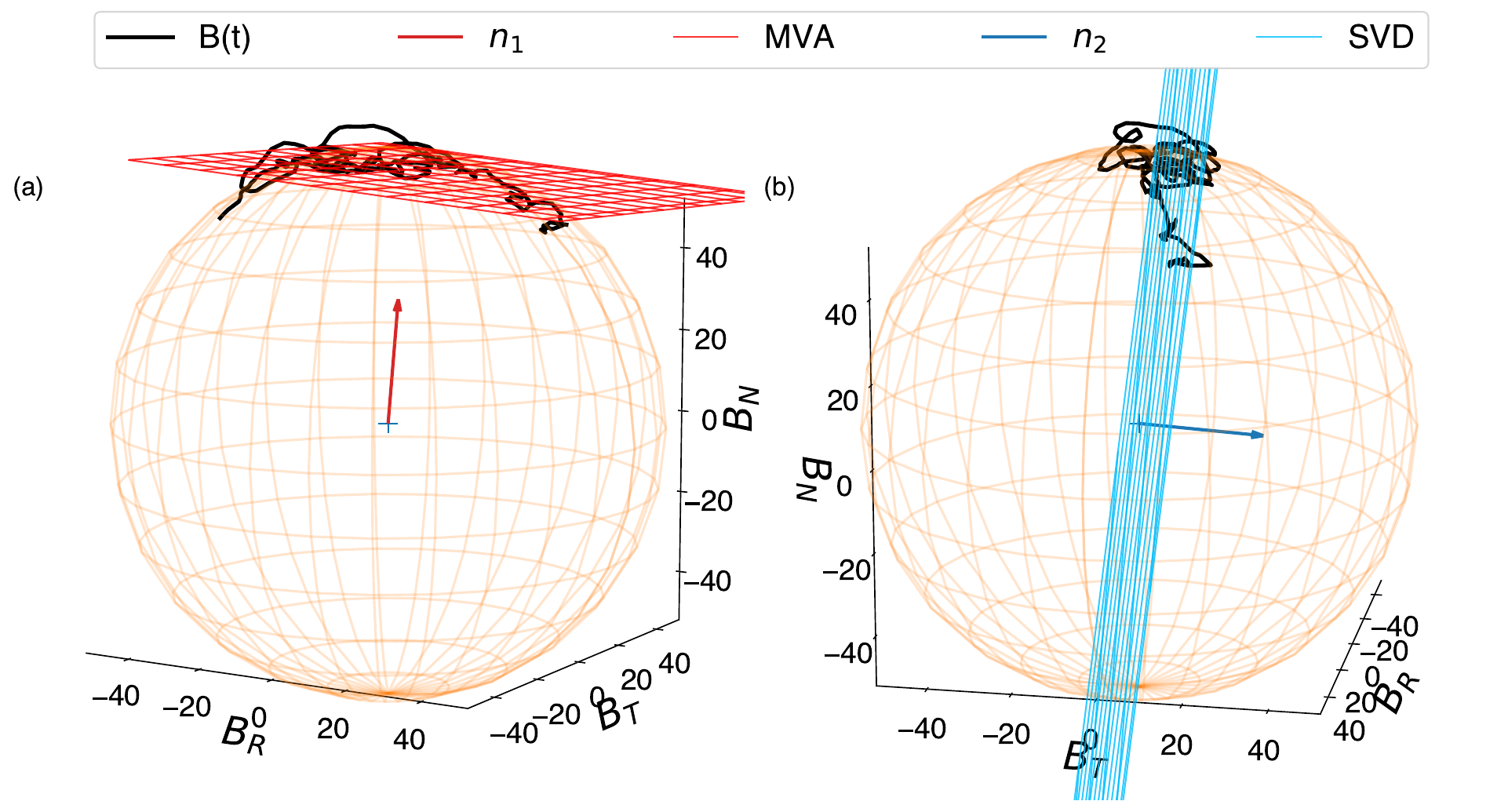}
\caption{Example of a typical switchback boundary in the RTN frame on 2018 November 7 at 20:53:15, which lasts $3 \si{s}$. The maximal deflection from the Parker spiral is about $104 \si{\degree}$. The black line corresponds to the measurements of the magnetic field during the selected boundary interval. The orange sphere, of radius $|\textbf{\textit{B}}|$, represents the (mostly) constant magnitude during the transition interval. The view is rotated between the two panels. Panel (a) shows in red the discontinuity plane identified by the MVA, with its normal $\textbf{\textit{n}}_{1}$ in red. Panel (b) shows in blue the discontinuity plane identified with the alternative method (SVD, see section~\ref{sec:svd}), with its normal $\textbf{\textit{n}}_{2}$ in blue. The animation has a real-time duration of 12 s and shows the rotation of the frame around the central axis of the sphere.\\
(An animation of this figure is available.)}
\label{fig_ex_132}
\end{figure*}

To illustrate how misleading the results of the MVA method can be, we represent in Figure~\ref{fig_ex_132} the orbit of the magnetic field for the leading edge of a switchback identified in the magnetic field waveforms, as obtained by MAG. For this particular case, we find $\lambda_1 / \lambda_2= 11.4$ and $\lambda_2 / \lambda_3= 2.1$. The classical selection criterion for normal identification is $\lambda_2 / \lambda_3 \ge 2$ \citep{leppingMagneticFieldDirectional1980}, although some authors recommend using a more stringent criterion $\lambda_2 / \lambda_3 \ge 10$ 
\citep{knetterFourpointDiscontinuityObservations2004}. 
Given the observed ratios, one might be tempted to conclude that the orbit is reasonably close to 2D and therefore the normal should coincide with the third principal axis. As it turns out, this conclusion is completely misleading, which leads us to look at the raw data.

Figure~\ref{fig_ex_132}a shows that the switchback boundary appears as an arc-shaped structure. Similar structures have been observed before in solar wind simulations \citep{vasquez96} and observations \citep{tsurutaniReviewAlfvenicTurbulence2018}, where they have been called arc-polarized structures. Notice how the arc is confined on the surface of a sphere $|\textbf{\textit{B}}|=$ const., which is a consequence of the high Alfvénicity of the solar wind. We are thus dealing with spherically polarized large-amplitude magnetic field fluctuations, in which switchback boundaries stand out as 2D structures that are plane polarized. These structures should be distinguished from the more usual case of plane polarization of finite-amplitude waves. The determination of the direction of minimum variance has been shown to be non-trivial for these spherically polarized structures \citep{barnesNonexistencePlanepolarizedLarge1976}.

\begin{figure}
    \centering
    \includegraphics[width=\hsize]{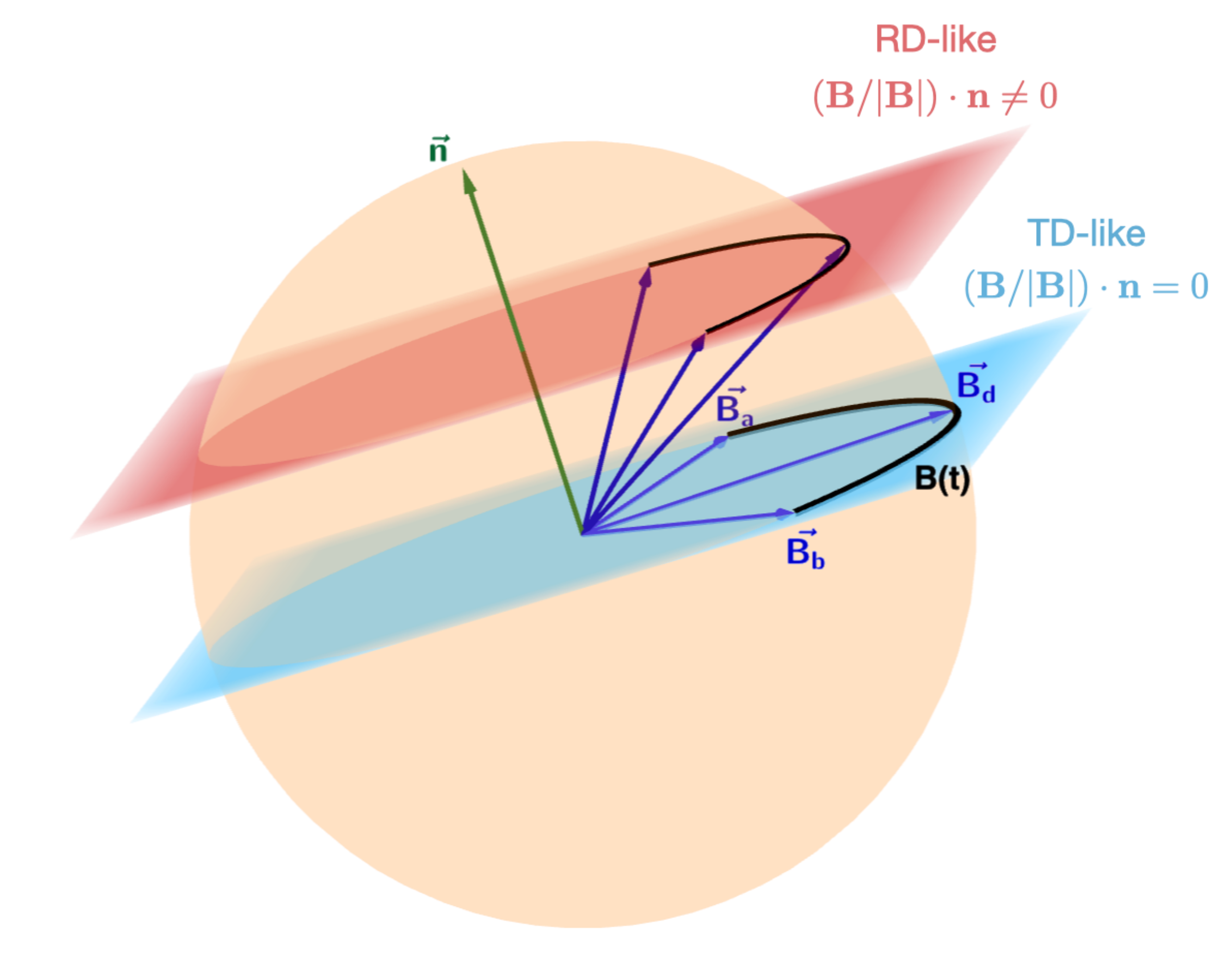}
\caption{Schematic view of an arc-polarized switchback boundary. The three vectors are from Equation (\ref{eq_normal}). The blue plane represents the discontinuity plane in which the tips of the field vector rotate, with its normal $\textbf{\textit{n}}$ in green. This is a particular event, for which the plane includes the origin. The orange sphere, of radius $|\textbf{\textit{B}}|$, represents the constant magnitude during the transition interval. A more general arc-polarized boundary is shown, with the identified plane in red. Both planes have the same normal. It should be noted that the two cases are ideal boundaries. Measured switchback boundaries would not have $(\textbf{\textit{B}}/|\textbf{\textit{B}}|) \cdot \textbf{\textit{n}} = 0$ exactly, mainly because of fluctuations.}
\label{fig_shema_rotation}
\end{figure}

There are two reasons why the MVA method shown in Figure~\ref{fig_ex_132}a fails to provide a correct estimate of the boundary normal. Surprisingly, these issues are known \citep{knetter05} and yet are often overlooked, especially in the analysis of switchback boundaries. First, as we shall show in the next section,  the orbit associated with the boundary always shows up as an arc-shaped structure, with fluctuations superimposed on it. Such an arc-shaped structure can be defined as the intersection between the sphere of constant $|\textbf{\textit{B}}|$ and a plane that cuts it, as illustrated in Figure~\ref{fig_shema_rotation}. From Equation~\ref{eq_normal}, we find the boundary normal to be perpendicular to this plane, i.e., essentially perpendicular to $\textbf{\textit{B}}$.

One would logically expect the MVA method to detect this plane because it is the only one that contains the orbit. What we find, however, is completely different: the MVA method detects a plane that is tangential to the surface of the sphere, shown in red in Figure~\ref{fig_ex_132}a. The boundary normal given by the MVA method points in the radial direction, and therefore tends to be aligned with the magnetic field. This situation prevails for most of the 125 switchbacks that we have analyzed.

A second reason for the ill-determination of the boundary normal has to do with fluctuations.  Switchback boundaries are affected by turbulence and surface waves \citep{krasnoselskikhLocalizedMagneticfieldStructures2020}, similar to those observed at boundaries farther away from the Sun \citep{horburyThreeSpacecraftObservations2001}. Such fluctuations cause the orbit to deviate from a regular arc to a fuzzy structure on a sphere. Because of the high Alfvénicity, the trajectory tends to remain on the surface of the sphere. When the deflection is small, this orbit defines a band-like structure, which the MVA method tends to identify as the plane of polarization, especially when the magnitude of the deflection is comparable to the level of fluctuations. This problem had already been pointed out by \citet{barnesInterplanetaryAlfvenicFluctuations1981}, who concluded that the direction of the minimum variance tends to be aligned with the magnetic field. This problem is exacerbated for small deflections, whose arc-shaped orbit can be difficult to disentangle from fluctuations. 

In conclusion, the automated application of the MVA method can easily produce boundary normals that are completely different from those obtained by visual inspection, leading to erroneous conclusions. For this reason, we will refrain from discussing the eigenvalues $\lambda_i$ as one would normally do, and instead focus on the geometric representation of the magnetic field vector. We cannot stress enough the importance of investigating the raw observations when determining the boundary normal. Let us now define a more robust method for determining the plane to which the orbit belongs.

\subsection{A different technique for plane fitting}
\label{sec:svd}

The problem of plane fitting in a high-dimensional space has received considerable attention for over a century \citep{pearson01} and is routinely solved by means of the singular value decomposition method (SVD) method \citep{golub13}. The starting point for these is a rectangular matrix that contains the three components of the magnetic field. The SVD method decomposes such a matrix into a product of three matrices: the first one contains the projection of the magnetic field on each of the three axes of the ellipsoid, the second one contains the values of $\sqrt{\lambda_i}$ and the third matrix contains the principal axes of the ellipsoid, one of which is $\textbf{\textit{n}}$.

The MVA method is conceptually identical to the SVD method, except that the mean value is subtracted from each of the components of the magnetic field prior to computing the SVD method (as in principal component analysis). By subtracting the mean, we allow the plane (when there is one) to be centered on the orbit. As discussed in the previous section, for switchback boundaries, the MVA method tends to find planes that are tangent to the sphere, as in Figure~\ref{fig_ex_132}a. 

If we do not subtract the mean (and thus apply the SVD method), then the plane is forced to include the origin, i.e. the center of the sphere $\textbf{\textit{B}} =0 $, see Figure~\ref{fig_ex_132}b. Therefore, by applying the SVD method to the raw data without first subtracting the mean, we force the plane to include the orbit of the magnetic field that describes an arc, shown in blue for an ideal boundary in Figure~\ref{fig_shema_rotation}. This is a special arc because the plane includes the origin, thus cutting the sphere in half. This arc defines the shortest path on the sphere between its two endpoints, similar to a geodesic. As we will see in the next section, the vast majority of switchback boundaries are remarkably well described by such arcs. For that reason, they are unique among all possible arc-polarized structures. However, for some structures, the plane clearly does not include the origin, as shown in red for an ideal boundary in Figure~\ref{fig_shema_rotation}, in which case we have  $(\textbf{\textit{B}}/|\textbf{\textit{B}}|) \cdot \textbf{\textit{n}} \neq 0$.

For switchbacks with small deflections and/or high levels of wave activity, the MVA method tends to find boundary normals aligned with the magnetic field. One may then incorrectly conclude that the discontinuity is RD-like. As the deflection increases and the arc structure becomes more apparent, the MVA solution becomes unstable. That is, the normal becomes increasingly sensitive to the choice of the time interval over which the mean of the magnetic field is computed. Eventually, for large deflections that typically exceed $90 \si{\degree}$, the MVA method switches solutions and detects the true plane of polarization rather than the one that was tangent to the sphere, as is the case for the example shown in Figure~\ref{fig_ex_30}.

In conclusion, the nature of switchback boundaries is closely related to a simple but specific geometric property: if the plane containing the discontinuity also contains the origin, then $\textbf{\textit{B}}$ and $\textbf{\textit{n}}$ are 
perpendicular and the boundary is TD-like. If it does not include the origin, then $\textbf{\textit{B}}$ and $\textbf{\textit{n}}$ are not
perpendicular and the boundary is RD-like. In this sense, determining how close the plane is to the origin is equivalent to specifying the nature of the boundary.  This property is emphasized by the two ideal boundaries, (i.e. without fluctuations) that are represented in Figure~\ref{fig_shema_rotation}. 

By carefully inspecting the orbit in 3D, and comparing the solutions obtained with/without subtracting the mean value of the magnetic field, we now have a combination of tools that gives us better insight into the topological properties of the orbit and should allow us to correctly determine the normal to the switchback boundary.

Note that this geometric property bypasses the need to specify the time interval for which the mean magnetic field must be determined for the MVA method, which has often been a contentious issue because of its impact on the determination of the normal \citep{hausmanDeterminingNatureOrientation2004}.

%
\section{Results}\label{sec_results}
\subsection{Orientation of the boundary normal and boundary classification}

The analysis of the MVA results in the previous section has shown that a blind application of this method can easily lead to erroneous conclusions. The problem goes beyond the usual one of relating the ratio of the variances $\lambda$ to the degree of planarity, which is often at the heart of discussions \citep{tsurutaniCommentCommentAbundances2007,tehLocalStructureDirectional2011,rosa_oliveira20}. As we have seen, even with seemingly reasonable ratios of the variances, the MVA method may completely fail to identify the proper boundary normal. For these reasons, it is essential to visually inspect each boundary to identify its nature.

From the visual inspection of 250 switchback boundaries that we selected during the first encounter of PSP, we observe that all behave as arc-polarized structures, which rotate on a sphere of constant $|\textbf{\textit{B}}|$ while also belonging to a plane that intersects this sphere. These structures are reminiscent of arc-polarized structures that have been observed in the solar wind near $1~\si{\astronomicalunit}$, and which are spherical in nature \citep{tsurutaniNonlinearElectromagneticWaves1997}. Because the boundaries are confined to a plane, their normal is well defined and can be estimated using Equation~\ref{eq_normal}. The 2D nature of these boundaries represents a strong topological constraint, whose origin has remained elusive so far, as in the case of arc-polarized structures that were observed in the solar wind \citep{tsurutaniReviewAlfvenicTurbulence2018}.

To determine the boundary normal, we use both the MVA and SVD methods and always visualize the structures in 3D. We consider the boundary normal to be well determined when at least one method gives a plane that includes the arc. Sometimes, both methods agree.
This is usually the case for switchbacks with large deflections. In most cases, however, the methods do not agree. When the plane containing the arc-shaped structure contains the origin, we rely only on the SVD method. Its solution does not necessarily give a normal that is exactly perpendicular to $\textbf{\textit{B}}$ because of wave activity. On the other hand, if the plane does not include the origin, then we estimate the normal using MVA and check the result visually. If the plane is too difficult to identify, then we discard the event for lack of clear evidence.

\begin{figure}
    \centering
    \includegraphics[width=\hsize]{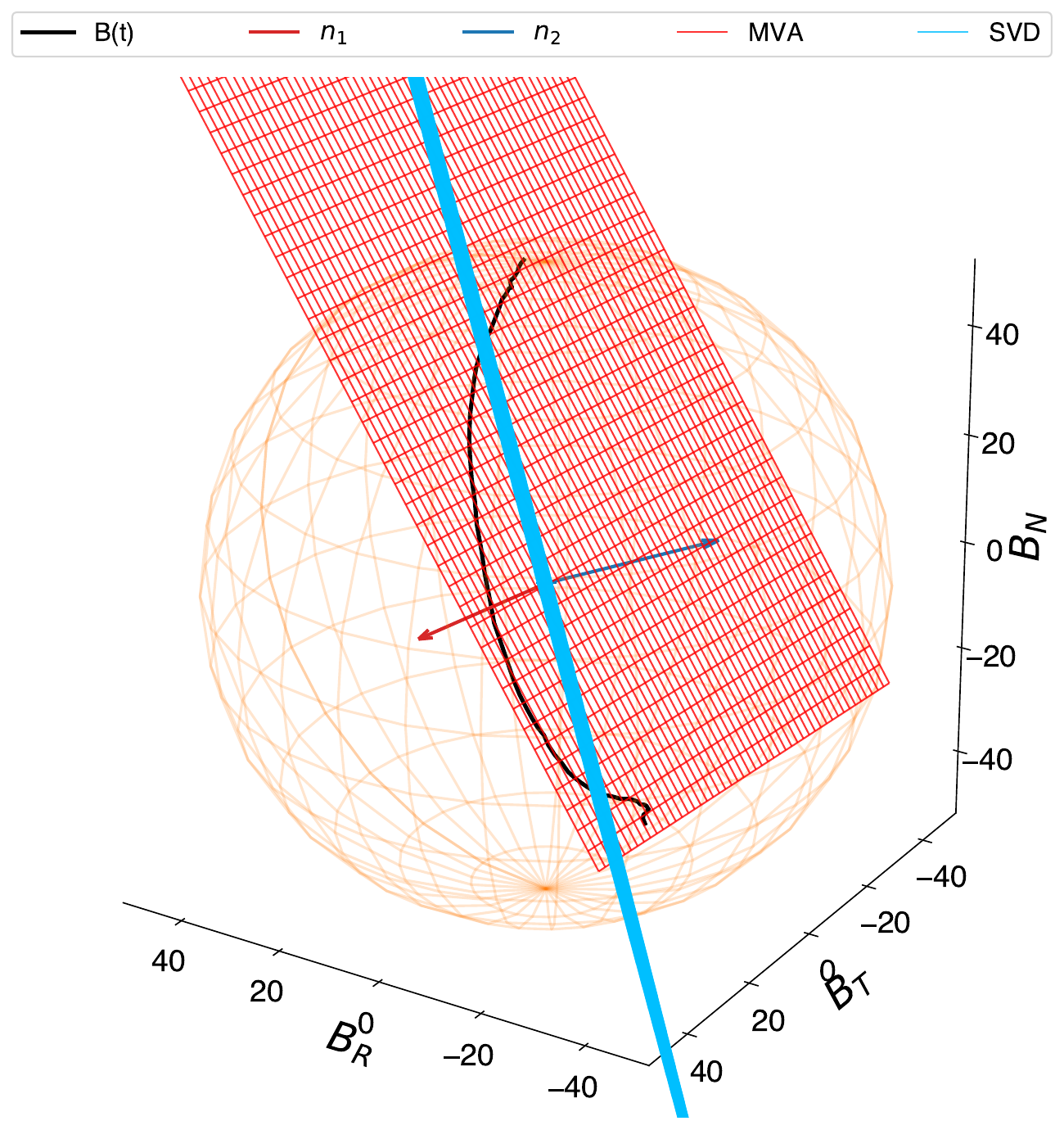}
    \caption{Same as Figure~\ref{fig_ex_132} but for a boundary misidentified by the SVD method but correctly identified by the MVA method. The event was detected on 2018 November 2 at
18:33:33, lasting for 2 s. The deflection from the Parker spiral is about $90 \si{\degree}$. The animation has a real-time duration of 12 s and shows the rotation of the frame around the central axis of the sphere.\\
(An animation of this figure is available.)}
    \label{fig_ex_30}
\end{figure}

Two examples of boundaries are shown in Figures~\ref{fig_ex_132} and \ref{fig_ex_30}. In the former, the MVA method identifies a boundary normal that is collinear with $\textbf{\textit{B}}$, which may therefore be mistaken for an RD. The SVD method better captures the arc (although this truly requires visualization in real 3D; animated versions of these figures are available in the online Journal and on Zenodo: doi:10.5281/zenodo.8424748), from which we conclude that the discontinuity is actually tangential.

Figure~\ref{fig_ex_30} illustrates one of the few cases in which the arc is well defined and yet the plane unambiguously excludes the origin. Here, the MVA method properly captures the structure, whereas the SVD method provides a wrong boundary normal.

Out of the 250 boundaries, 186 (74 \%) are correctly classified by our methodology, 179 (71 \%) are located in a plane that includes the origin, seven (3\%) have a plane that clearly does not include the origin, and 64 (26 \%) are in between. The latter generally correspond to switchbacks with small deflections, whose boundary is blurred by wave activity and turbulence.

\subsection{Comparison with classical boundary classification and previous studies}

To allow for a comparison between our results and previous studies, we investigate the nature of each boundary by using the classical single-spacecraft classification scheme for boundaries \citep{smithIdentificationInterplanetaryTangential1973}. This scheme relies on two quantities: the degree of collinearity $(\textbf{\textit{B}}/|\textbf{\textit{B}}|) \cdot \textbf{\textit{n}}$, and the change in field magnitude across the boundary $\Delta |\textbf{\textit{B}}|/|\textbf{\textit{B}}|$. 
\cite{neugebauerReexaminationRotationalTangential1984} adopted a threshold value of 0.4 for the former and 0.2 for the latter to clearly distinguish TDs from RDs. These criteria are still widely used today in the literature \citep{tsurutaniInterplanetaryDiscontinuitiesAlfven1996, horburyThreeSpacecraftObservations2001}. Note that a high collinearity would provide unambiguous evidence for an RD whereas small values can imply either TDs or RDs \citep{knetter05}.

\begin{figure}
\resizebox{\hsize}{!}{\includegraphics{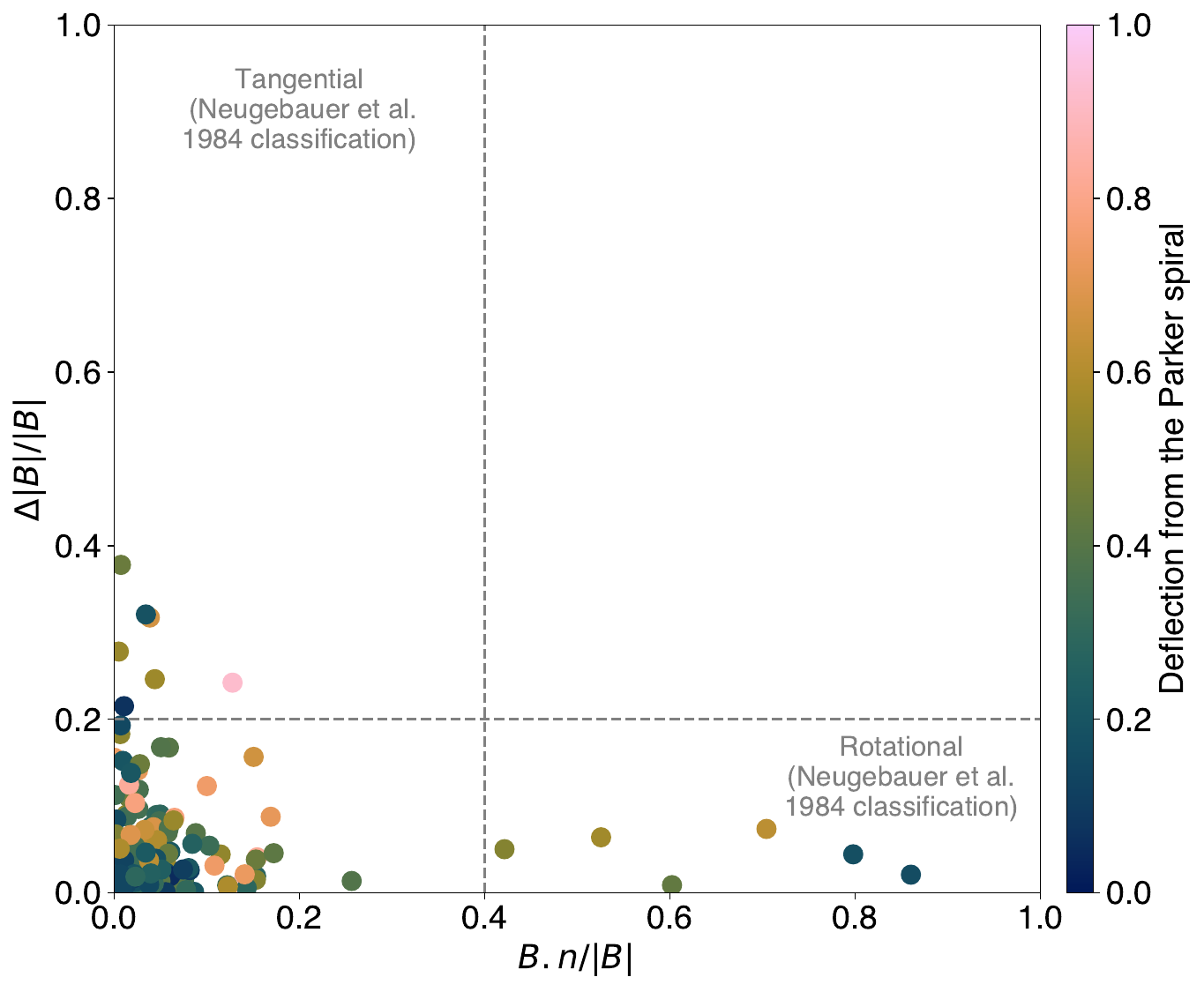}}
\caption{ 
Boundary analysis using the classical classification scheme by \cite{neugebauerReexaminationRotationalTangential1984}. The normalized projection of the magnetic field along the normal $(\textbf{\textit{B}}/|\textbf{\textit{B}}|) \cdot \textbf{\textit{n}}$ is plotted against the variation of the magnetic field magnitude across the discontinuity $\Delta |\textbf{\textit{B}}|/|\textbf{\textit{B}}|$. The dashed lines divide the plane into four quadrants: TDs occur in the upper left quadrant and RDs in the lower right one. The remaining quadrants correspond to unclassified events. The color code represents the normalized magnitude $z$ of the deflection from the Parker spiral (Equation~(\ref{eq_deflection})).
186 boundaries are shown, excluding the 64 ambiguous cases.
}
\label{fig_ex_results_stat}
\end{figure}

Figure \ref{fig_ex_results_stat} shows how our boundaries are distributed, using the two-parameter classification scheme of \cite{neugebauerReexaminationRotationalTangential1984}. We only show those boundaries for which the normal can be clearly identified  with our methodology, i.e. 186 cases out of 250.

This figure leads to several important results. First, we find that few boundaries qualify as pure TDs or RDs. Most boundaries belong to the so-called limit category with $(\textbf{\textit{B}}/|\textbf{\textit{B}}|) \cdot \textbf{\textit{n}} \rightarrow 0 $, for which TDs and RDs cannot be distinguished based only on magnetic field data.

To overcome this ambiguity between both types of discontinuities, we examine the necessary conditions for having perfect RDs or TDs. For classical TDs, we expect an arbitrarily large jump in the magnitude of the magnetic field. However, this is not a necessary condition in our context: when the structures are Alfvénic and the change in the magnetic field at their boundary is modest, the jump can be arbitrarily small while the discontinuity still remains TD-like. This is precisely the case for switchbacks, which are known to be highly Alfvénic \citep{kasperAlfvenicVelocitySpikes2019, lakerSwitchbackDeflectionsEarly2022}.

Additional conditions are linked to the velocity jumps at the boundary.
A necessary one for having RDs is a large value of the ratio $R_{\textit{VB}}$ between the change across the boundary in the tangential component of the velocity ${v}_t$ and that of the magnetic field ${B}_{t}$ \citep{baumjohannBasicSpacePlasma2012}
\begin{equation}\label{eq_ratio_rvb}
R_{\textit{VB}}=\sqrt{\mu_{0}\rho}\frac{[{v}_t]}{[{B}_{t}]},
\end{equation}
where $\rho$ is the proton density at the discontinuity. The closer the ratio is to 1, the more likely we identified RDs, showing the correlated changes between the tangential components to the velocity and the magnetic field. On the contrary, it can have any value for TDs, which are expected to be closer to 0.

We calculate this ratio on a subset of 42 discontinuities, which presents enough and good velocity measurements. We obtain a broad distribution of ratios, with $85\%$ of the ratios with values lower than 0.2, thus clearly excluding the possibility of having RDs. In addition, we do not find a significant correlation between the fluctuations of ${v}_t$ and ${B}_t$, as would be expected for RDs.

Taken together, these different results all point to the tangential rather than rotational nature of the discontinuities. The boundaries located in the lower left corner of Figure \ref{fig_ex_results_stat} are therefore more likely to be TDs because of the small value of $(\textbf{\textit{B}}/|\textbf{\textit{B}}|) \cdot \textbf{\textit{n}}$, even though the relative change in $|\textbf{\textit{B}}|$ is smaller than what is usually encountered with (mostly non-Alfvénic) discontinuities. We will refer to them as TD-like. Only multipoint measurements could help us fully resolve this ambiguity. An additional source of uncertainty comes from the slow cadence of the proton instruments compared to that of the magnetometer, which prevents us from properly resolving the details of most boundaries.

Now that we know that the vast majority of discontinuities are TD-like, a second important finding is the stark contrast between our conclusion and that of previous studies that considered the same time interval.  Indeed, \citet{larosaSwitchbacksStatisticalProperties2021} found that RDs are slightly more common than TDs, while \citet{akhavan-taftiDiscontinuityAnalysisLeading2021} argued that $78\%$ of the boundaries are RDs. We now know that these earlier studies most likely overestimated the number of RDs because of the automatic application of the MVA method. Most discontinuities are actually much closer to tangential, and therefore the switchback boundary is more impermeable to plasma flow than expected. This has important implications for the rate at which switchbacks can erode in the solar wind.

A third result deals with the magnitude of the deflection of the switchbacks. Figure~\ref{fig_ex_results_stat} includes deflections of all magnitudes, including those that strictly qualify as switchbacks (for which $z>0.5$). Interestingly, the type of discontinuity does not depend on the magnitude of the deflection: small deflections behave similarly to large deflections. This confirms earlier observations by \citet{dudokdewitSwitchbacksNearSunMagnetic2020} that these deflections are self-similar, with no characteristic cutoff beyond which they behave differently. This is true down to a value of approximately $z=0.1$, below which it becomes increasingly difficult to distinguish switchback boundaries from random fluctuations. These results suggest that small deflections are unlikely to be the result of an erosion of large reversal structures.

Among the processes that could erode these boundaries is reconnection, which has been observed on a limited number of boundaries. \citet{fromentDirectEvidenceMagnetic2021}, for example, showed three examples of boundaries with evidence of reconnection. However, we do not see a causal link between the few identified RDs and the few cases of reconnection at the switchback boundary. Actually, the reconnection events from \citet{fromentDirectEvidenceMagnetic2021} are identified as TDs with our methodology. 

Moreover, as we find more numerous TDs than RDs, the consequent survival of the structures during their outward propagation is compatible with an origin that is rooted in the solar atmosphere. However, our results do not provide a sufficient condition for excluding the in situ generation of switchbacks.

\subsection{Thickness of the boundaries}

Now that we have more realistic estimates of the switchback boundary normals, we can estimate the thickness of these boundaries. Indeed, these estimates depend heavily on the orientation of the normals. We infer the thickness of the boundaries from the normal velocity of the protons and the duration of the boundaries in the MAG time series. For some events, the proton velocity could not be used, so our sample contains 165 events out of the 186 classified boundaries. The classified boundaries correspond to those identified by our method.

We determine the thickness of each boundary from
\begin{equation}\label{eq_thickness}
D=\left ( t_{d}-t_{b} \right ) \cdot \left( \langle\textbf{V}_{pl}\rangle-\textbf{V}_{SC}\right) , 
\end{equation} 
where $D$ is the thickness of the boundary, $t_{b}$ and $t_{d}$ the time in seconds of the beginning and end of the boundaries in the time series, respectively; $\langle\textbf{V}_{\textit{pl}}\rangle$ is the mean plasma velocity outside the switchback and $\textbf{V}_{\textit{SC}}$ is the spacecraft velocity. Both velocities are in the RTN frame projected in the normal direction. 

Theoretically, the propagation velocity of the discontinuity with respect to the spacecraft should be included in the calculation. However, we cannot determine it with a single spacecraft, and so the estimation of the thickness relies on the assumption of the nature of the discontinuity to include or not the Alfvén velocity, for TDs and RDs, respectively \citep{knetter05}.

\begin{figure}
\resizebox{\hsize}{!}{\includegraphics{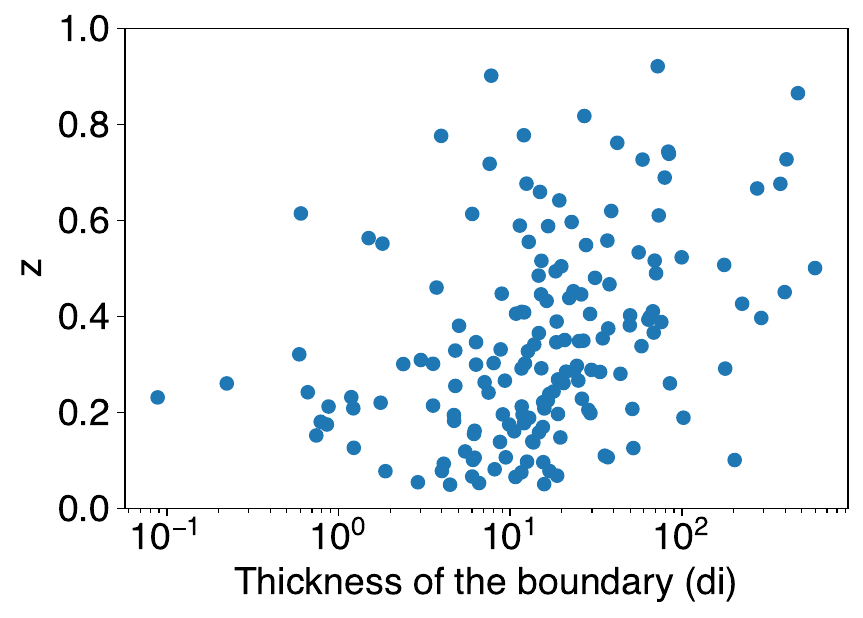}}
\caption{Boundary thickness as a function of the magnitude of their deflection from the Parker spiral. The thickness is calculated in ion inertial length. The mean ion inertial length over the first encounter of PSP is $16 \si{km}$.}
\label{fig_ex_results_thickness}
\end{figure}

Figure \ref{fig_ex_results_thickness} shows the thickness of the 165 boundaries as a function of the magnitude of the deflection from the Parker spiral. First, we note that the distribution of thicknesses is a factor of 3-10 smaller than those found by \citet{larosaSwitchbacksStatisticalProperties2021}. This difference was to be expected because we provide better estimates of the normals, which are more perpendicular to $\textbf{\textit{B}}$ than the ones found by \citet{larosaSwitchbacksStatisticalProperties2021}.

Interestingly, most of the thicknesses exceed the proton inertial length, so the boundaries can be considered MHD discontinuities. We also find that the magnitude of the deflection has no effect on the thickness of the boundary. This supports the self-similar geometry we alluded to earlier, where all switchbacks belong to the same population of structures, regardless of the magnitude of their deflection. Third, we find no clear correlation between the thickness and the type of boundary. In addition, we do not observe differences in thickness between the leading and trailing edges when considering the same structures, and no clear asymmetry is observed in the structures depending on their position at the first encounter, such as corotation times or longitudinal sweep.

\section{Conclusions} \label{sec_conclusion}

In conclusion, we have investigated different methods for studying the boundary of magnetic switchbacks as observed by PSP during its first solar encounter. We find that these boundaries have a very specific geometrical shape, since all are arc-polarized structures whose rotation is contained in a plane. 

Our findings also caution against the routine use of MVA for identifying the nature of boundaries of structures that are highly Alfvénic, regardless of the ratio of its eigenvalues. We find that the presence of fluctuations often causes the method to wrongly identify boundary normals that are collinear with the magnetic field, i.e. it has a bias toward boundaries that appear to be rotational. We thus propose a rigorous methodology to identify the nature of the switchback boundaries, by a better estimation of the normal.

A large majority (71\%) of these planes include the origin. This specific geometry implies that the normal is mostly perpendicular to the magnetic field. Very few (3\%) of the planes do not include the origin and thus qualify as rotational discontinuities. The remainder (23\%) is unclassified by lack of clear orientation of the normal. 

Although we cannot provide definite proof of the nature of these discontinuities, there is considerable evidence to suggest that most of them are not rotational, and therefore tangential-like. This conclusion is in contrast to previous analyses that found a predominance of rotational discontinuities. In turn, it suggests that switchback boundaries are more likely to be closed. These conclusions raise another aspect that should be considered in future studies. A tangential discontinuity does not have a velocity relative to the plasma, so it does not propagate in the solar wind. Thus, the velocity should not have an influence on the switchback crossing. This is contradicted by some of the observations, which show small but noticeable velocity spikes.
 
Our analysis further reveals no clear influence of the magnitude of the deflection from the Parker spiral on the type of boundary nor on their thickness, confirming the self-similar nature of switchbacks.

Taken together, our results challenge the classical picture of switchbacks and suggest their erosion could be much slower than expected. Switchbacks, with their stable boundaries, could survive longer as they propagate out into the solar wind.

\begin{acknowledgments}
The FIELDS experiment was developed and is operated under NASA contract NNN06AA01C. N.B., T.D., and C.F. acknowledge financial support of the CNES in the frame of a Parker Solar Probe grant. N.B. and T.D. thank Jaye Verniero for stimulating discussions. N.B., T.D., and C.F. thank Vladimir Krasnoselskikh for the interesting comments. Parker Solar Probe was designed, built, and is now operated by the Johns Hopkins Applied Physics Laboratory as part of NASA’s Living with a Star (LWS) program (contract NNN06AA01C). Support from the LWS management and technical team has played a critical role in the success of the Parker Solar Probe mission.
The data used in this study are available at the NASA Space Physics Data Facility (SPDF),  \href{https://spdf.gsfc.nasa.gov}{https://spdf.gsfc.nasa.gov}.
\end{acknowledgments}

\bibliography{sample631}{}
\bibliographystyle{aasjournal}

\end{document}